\documentclass[aps,preprintnumbers,showpacs,showkeys,nofootinbib,
superscriptaddress,fleqn,floatfix,tightenlines,preprint]{revtex4-1}
\usepackage{amsmath,amsfonts,amssymb,amscd,amsxtra,amsthm}
\usepackage{graphicx,slashed,bm}
\usepackage{mathrsfs,multirow,booktabs,epstopdf}
\usepackage[latin9]{inputenc}
\usepackage[normalem]{ulem} 
\usepackage[dvipsnames]{xcolor} 
\renewcommand\sout{\bgroup \color{red} \ULdepth=-.5ex \ULset}

\begin{document}
\preprint{INHA-NTG-05/2016}
\title{Heavy pentaquark states $P_c(4380)$ and $P_c(4450)$ in the
  $J/\psi$ production induced by pion beams off the nucleon}
\author{Sang-Ho Kim}
\email[E-mail: ]{sangho.kim@apctp.org}
\affiliation{Asia Pacific Center for Theoretical Physics (APCTP), Pohang,
Gyeongbuk, 790-784, Republic of Korea}
\author{Hyun-Chul Kim}
\email[E-mail: ]{hchkim@inha.ac.kr}
\affiliation{Department of Physics, Inha University, Incheon 402-751,
Republic of Korea}
\affiliation{School of Physics, Korea Institute for Advanced Study
  (KIAS), Seoul 130-722, Republic of Korea}
\author{Atsushi Hosaka}
\email[E-mail: ]{hosaka@rcnp.osaka-u.ac.jp}
\affiliation{Research Center for Nuclear Physics (RCNP), 
Osaka University, Ibaraki, Osaka, 567-0047, Japan}
\affiliation{J-PARC Branch, KEK Theory Center, 
Institute of Particle and Nuclear Studies, KEK, 
Tokai, Ibaraki, 319-1106, Japan}
\date{\today}
\begin{abstract}
In this study, we investigate the $J/\psi$ production induced by pion
beams off the nucleon, particularly the heavy pentaquarks $P_c(4380)$
and $P_c(4450)$ in intermediate states, based on a hybridized Regge
model. The process involving $\rho$ and $\pi$ meson exchange in the
$t$ channel is considered as background, and the heavy pentaquark 
exchange is included in the $s$ channel. The coupling constants such
as the $\rho NN$ and $\pi NN$ vertices are taken from the $NN$
potentials, whereas those for the $J/\psi\rho\pi$ and $J/\psi\pi\pi$
vertices are determined by using experimental data based on the
branching ratios. In order to estimate the $P_c(4380)$ and $P_c(4450)$
coupling constants, we use the experimental upper limit on
the total cross section as a guide for the $\pi N\to J/\psi N$ reaction. The
background total cross section is the order of $10^{-4}-10^{-3}$
nb. In the vicinity of the heavy pentaquark masses, the total cross
section reaches about $1$ nb. 
\end{abstract}
\maketitle
\section{Introduction}
\label{SecI} 
Finding new exotic hadrons is one of the most important issues for
hadron and particle physics. Recently, the LHCb Collaboration 
announced the observation of two heavy pentaquarks in $\Lambda_b\to
J/\psi K^- p$ decays~\cite{Aaij:2015tga}, where the quark content 
is $uudc \bar c$. The significance of these pentaquarks is more than
9$\sigma$. The masses and widths were reported as: $M_{P_c}=(4380 \pm 8
\pm 29)$ MeV and $\Gamma_{P_c}= (205 \pm 18 \pm 86)$ MeV for the lower
state, whereas $M_{P_c}=(4449.8 \pm 1.7 \pm 2.5)$ MeV and
$\Gamma_{P_c}=(39 \pm 5 \pm 19)$ MeV for the higher state.
Thus, it is very important to confirm these pentaquark states in other
possible reactions. For example, the energy of the pion beam at the
Japan Proton Accelerator Research Complex (J-PARC) facility is 
sufficient to observe them during the $\pi N\to J/\psi N$ process. The
photon beam at the Thomas Jefferson National  Accelerator Facility
(Jefferson Lab) can also be used to measure the $P_c$ states in $J/\psi$ 
photoproduction~\cite{Wang:2015jsa,Kubarovsky:2015aaa,Karliner:2015voa}.

In fact, a recent theoretical study of the $\pi^- p \to
J/\psi n$ reaction with neutral charm pentaquarks $P_c^0$
~\cite{Lu:2015fva} employed the effective Lagrangian method. Lu et
al.~\cite{Lu:2015fva} assumed that the branching ratios of $P_c\to
J/\psi N$ and $P_c\to \pi N$ 
are about 10~\% and 1~\%,
respectively. The background total cross section is in the order of $10-100$
nb and the total cross section near the pentaquark masses increases
to around $1\,\mathrm{\mu b}$. This indicates that the magnitude of
the total cross section for $J/\psi$ production in the vicinity of
the pentaquark masses is almost comparable to that of the $\pi N\to
\phi N$ reaction. 

\begin{figure}[hb]
\centering
\includegraphics[scale=0.55]{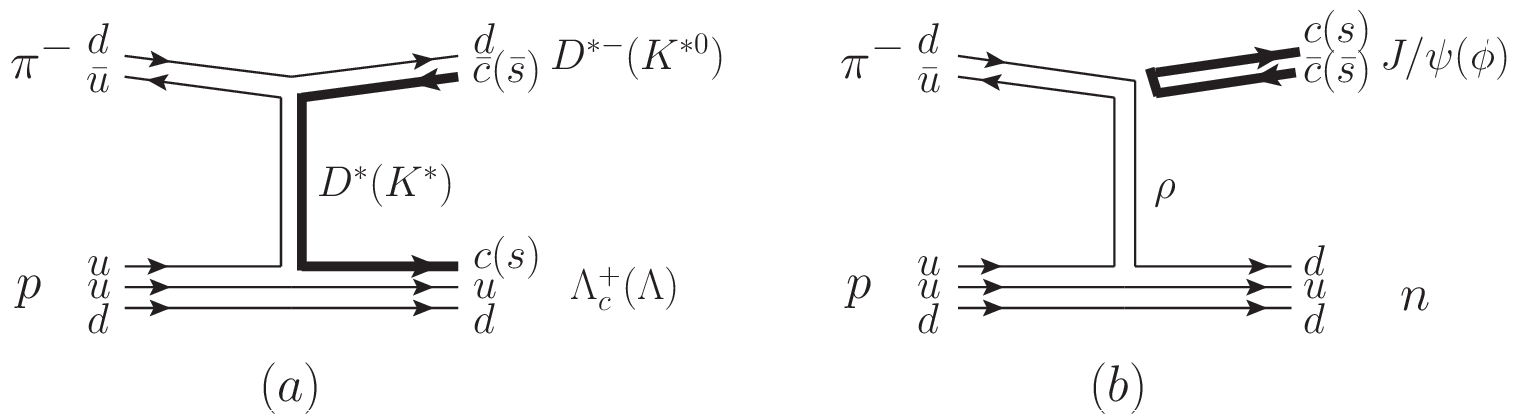}
\caption{Quark diagrams for (a) open and (b) hidden charm 
(strangeness) production.
(a) $\pi^- p \to D^{*-} \Lambda_c^+ (K^{*0} \Lambda)$ and
(b) $\pi^- p \to J/\psi n (\phi n)$.}  
\label{FIG1}
\end{figure}

In the present study, we consider the contribution
of pentaquark resonances to $J/\psi$ production by using  
previous experimental information on the upper limit of the $\pi N \to
J/\psi N$ reaction~\cite{Jenkins:1977xb, Chiang:1986gn} and by employing 
a hybridized Regge model that incorporates the heavy pentaquark states. 
To estimate the contribution of the heavy pentaquarks to the $\pi N\to
J/\psi N$ reaction, it is crucial to know the background
contribution. Recently, the open charm production $\pi^- p \to
D^{*-}\Lambda_c^+$ was analyzed based on a comparison with the 
associated strangeness production 
$\pi^- p \to K^{*0}\Lambda$~\cite{Kim:2014qha,Kim:2015ita,e50}
(see Fig.~\ref{FIG1}(a)), and thus the parameters for the $\pi^- p
\to D^{*-}\Lambda_c^+$ reaction can be plausibly estimated. A similar
approach was also applied to various reactions such as  
$\bar p p \to \bar Y_c Y_c$ and $\bar p p \to \bar M_c M_c$, where
$Y_c$ and $M_c$ denote $\Lambda_c^+, \Sigma_c^+$ and $D,
D^*$~\cite{Titov:2008yf}. 

We adopt the same strategy to study the hidden charm process $\pi^- 
p \to J/\Psi n$ together with the strangeness process $\pi^- p \to
\phi n$ (see Fig.~\ref{FIG1}(b)). However, the hidden charm
(strangeness) reactions are distinguished from the open charm
(strangeness) reactions. In the case of the open charm processes, the
exchanged meson in the $t$ channel should be different from that in
the open strangeness process, as shown in Fig.~\ref{FIG1}(a). In
addition, $\rho$ meson and pion exchanges play similar roles in
both the hidden charm and strangeness reactions because of the 
Okubo-Zweig-Iizuka (OZI) suppression, which is illustrated in
Fig.~\ref{FIG1}(b). 
This allows us to obtain the coupling constants more explicitly by using
the corresponding experimental data rather than relying on 
model calculations.  

This Letter is organized as follows. In Section~\ref{SecII}, we
explain the general formalism of a hybridized Regge model and we show how
to determine the relevant parameters for the $\pi$ and $\rho$ Reggeons
and the $P_c$ pentaquark states. In Section~\ref{SecIII}, we
first examine the assumption made in previous studies regarding the
branching ratios of $P_c\to J/\psi N$ and $P_c \to \pi N$, where we
compute the total cross section for the $\pi N\to J/\psi N$. We also
present numerical results for the total cross-section and differential
cross sections. In Section~\ref{SecIV}, we give our conclusions and a
summary of the present study.

\section{General formalism}
\label{SecII}
\begin{figure}[ht]
\centering
\includegraphics[scale=0.65]{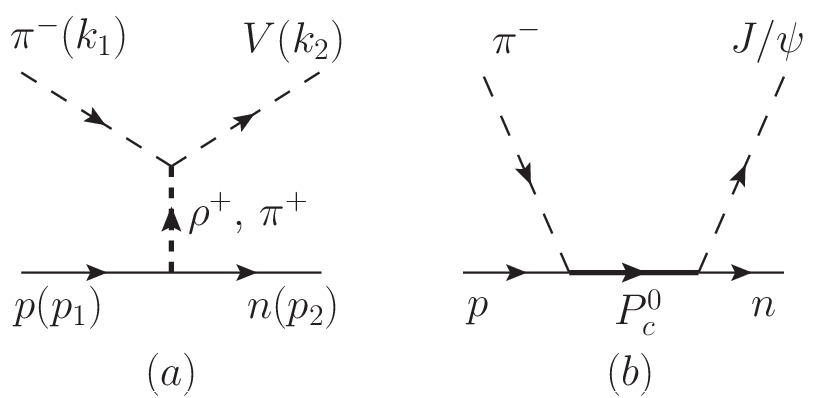}
\caption{Tree-level Feynman diagrams for 
(a) $\pi^- p \to V (\phi,J/\psi) n$ in the $t$ channel and 
(b) $\pi^- p \to J/\psi n$ in the $s$ channel.}
\label{FIG2}
\end{figure}

In this study, we employ a hybridized Regge model for the charm production 
$\pi^- p \to J/\psi n$ in order to consider the contribution of the charm 
pentaquark states $P_c^0$ together with the strangeness production  
$\pi^- p \to \phi n$. Figure~\ref{FIG2}(a) shows a generic $t$-channel tree
diagram for $\pi^- p \to V (\phi,J/\psi) n$, and 
Fig.~\ref{FIG2}(b) depicts $P_c^0$ exchange in the $s$ channel only
for the $\pi^- p \to J/\psi n$ process. The initial momenta of the
pion and the proton are denoted by $k_1$ and $p_1$, respectively, and
the final momenta of the vector meson and the neutron are denoted
by $k_2$ and $p_2$, respectively.

The effective Lagrangians for the exchanges of the $\rho$ and $\pi$ mesons
are expressed as  
\begin{align}
\mathcal L_{V\rho\pi} &= \frac{g_{V\rho\pi}}{M_V}
\varepsilon^{\mu\nu\alpha\beta}
\partial_\mu V_\nu \partial_\alpha \bm{\rho}_\beta \cdot\bm{\pi},
\label{eq:Lag1}         \\ 
\mathcal L_{V\pi\pi} &= 
-ig_{V\pi\pi} (\pi^- \partial_\mu \pi^+ - \partial_\mu \pi^- \pi^+) V^\mu,
\label{eq:Lag2} 
\end{align}
where $V = (\phi,\,J/\psi)$ and $\pi,\,\rho,\,\phi$, and $J/\psi$
denote the fields corresponding to the
$\pi(140,0^-)$, $\rho(770,1^-)$, $\phi(1020,1^-)$, and $J/\psi(3097,1^-)$ 
mesons, respectively.  
The coupling constants for $\rho$ exchange are determined by using the
decay width of the corresponding vector mesons~\cite{Olive:2014zz}
\begin{align}
\Gamma (V \to \rho \pi) 
= \frac{1}{4\pi} \frac{1}{M_V^2} |\vec k|^3 g_{V\rho\pi}^2,
\label{eq:CC:VRhoPi}
\end{align}
where $|\vec k| =
\sqrt{[M_V^2-(M_\rho+M_\pi)^2][M_V^2-(M_\rho-M_\pi)^2]}/(2M_V)$.
Similarly, the coupling constants $g_{V\pi\pi}$ are derived 
from experimental data on the decay width~\cite{Olive:2014zz}
\begin{align}
\Gamma (V \to \pi^+ \pi^-) 
= \frac{1}{6\pi} \frac{1}{M_V^2} |\vec q|^3 g_{V\pi\pi}^2,
\label{eq:CC:VPiPi}
\end{align}
where $|\vec q| = \sqrt{M_V^2-4M_\pi^2}/2$. 
All of the relevant numerical values for the couplings of $\phi$ and 
$J/\psi$ mesons~\cite{Olive:2014zz} are summarized in Table~\ref{TABLE1}.
In fact, the Particle Data Group provides the branching ratio 
$\Gamma(\phi\to(\rho\pi + \pi^+\pi^-\pi^0))/\Gamma(\phi) = 15.32\,\%$
~\cite{Olive:2014zz}. 
However, we ignore this part of the 3$\pi$ decay in the data, where we assume 
that it is smaller than the $\rho \pi$ channel. The $\rho$ meson 
has a rather large width, but we find that its effect is very small, as stated
previously~\cite{Sibirtsev:1998cs}. Hence, we neglect the effect 
of the $\rho$ meson decay width when calculating the coupling constant
$g_{V\rho\pi}$. 
\begin{table}[hptb]
\begin{tabular}{cc||cc|c||cc|c} \hline\hline
$V$
&$\Gamma_V[\mathrm{MeV}]$
&$\Gamma_{V\to\rho\pi}/\Gamma_V[\%]$
&$\Gamma_{V\to\rho\pi}[\mathrm{keV}]$
&$g_{V\rho\pi}$
&$\Gamma_{V\to\pi^+\pi^-}/\Gamma_V[\%]$
&$\Gamma_{V\to\pi^+\pi^-}[\mathrm{eV}]$
&$g_{V\pi\pi}$ \\ \hline
$\phi$
&$4.266$
&$15.32 \pm 0.32$
&$654$
&$1.25$
&$(7.4 \pm 1.3) \cdot 10^{-3}$
&$316$
&$7.24 \cdot 10^{-3}$ \\$J/\psi$
&$0.0929$
&$1.69 \pm 0.15$
&$1.57$
&$7.90 \cdot 10^{-3}$
&$(1.47 \pm 0.14) \cdot 10^{-2}$
&$13.7$
&$8.20 \cdot 10^{-4}$ \\\hline\hline
\end{tabular}
\caption{Coupling constants $g_{V\rho\pi}$ and $g_{V\pi\pi}$,
  based on the
branching ratios of $V(\phi,J/\psi)$ relative to $\rho\pi$ and to  
$\pi^+\pi^-$~\cite{Olive:2014zz}.} 
\label{TABLE1}
\end{table}

The effective Lagrangians for the $\rho NN$ and $\pi NN$ vertices are as follows
\begin{align}
\mathcal L_{\rho N N} &= 
-g_{\rho N N} \left[ \bar N \gamma_\mu \bm{\tau} N - 
\frac{\kappa_{\rho N N}}{2 M_N} \bar N \sigma_{\mu\nu} \bm{\tau} N \partial^\nu 
\right] \cdot \bm{\rho}^\mu,                                          \cr  
\mathcal L_{\pi N N} &=
-i g_{\pi N N} \bar N \gamma_5 \bm{\tau}\cdot\bm{\pi} N,             
\label{eq:Lag3}
\end{align}
where $N$ denotes the nucleon and the coupling constants are taken
from the $NN$ potentials, i.e., $g_{\rho N N}=3.36$, $\kappa_{\rho N
  N}=6.1$ and $g_{\pi N N}=13.3$ (e.g., see ~\cite{Machleidt:1987hj}). 

The invariant amplitudes are derived in the form of 
\begin{align}
\mathcal{M} = \bar{u}_n \mathcal{M}^\mu \epsilon_\mu^* u_p,
\end{align}
where 
\begin{align}
\mathcal M_\rho^\mu &= I_\rho
\frac{g_{V\rho\pi} g_{\rho N N}}{t-M_\rho^2}
\epsilon^{\mu\nu\alpha\beta} 
\left [ \gamma_\nu - \frac{i\kappa_{\rho N N}}{2M_N} 
\sigma_{\nu \lambda}(k_2-k_1)^\lambda \right ] k_{2 \alpha} k_{1 \beta}, \cr  
\mathcal M_\pi^\mu &= -2I_\pi
\frac{ig_{V\pi\pi}g_{\pi N N}}{t-M_\pi^2} \gamma_5  k_1^\mu.
\label{eq:InvAmp}                     
\end{align}
$u_p$ and $u_n $ denote the Dirac spinors of the incoming and 
outcoming nucleons, respectively, and $\epsilon_\mu$ is the polarization 
vector of the final vector meson. The isospin factors are given by
$I_\rho = I_\pi = \sqrt{2}$.  

We now replace the Feynman propagators in Eq.~(\ref{eq:InvAmp}) with the
Regge propagators $R(t)$ as~\cite{Guidal:1997hy}
\begin{align}
\frac{1}{t-M_\rho^2} &\to
R_\rho(t) = \left( \frac{s}{s_\rho} \right)^{\alpha_\rho(t)-1}
\frac{1}{\mathrm{sin}[\pi\alpha_\rho(t)]}
\frac{\pi\alpha_\rho'}{\Gamma[\alpha_\rho(t)]},                        \cr  
\frac{1}{t-M_\pi^2} &\to 
R_\pi(t) = \left( \frac{s}{s_\pi} \right)^{\alpha_\pi(t)}
\frac{1}{\mathrm{sin}[\pi\alpha_\pi(t)]}
\frac{\pi\alpha_\pi'}{\Gamma[1+\alpha_\pi(t)]},
\end{align}
where the Regge trajectories are given by $\alpha_\rho(t) = 0.55 +0.8t$
and $\alpha_\pi(t) = 0.7(t - M_\pi^2)$~\cite{Guidal:1997hy}.
The energy scale parameters are 
selected as $s_\rho=s_\pi=1\,\mathrm{GeV^2}$ for simplicity.
For the signature factor, we select a
constant phase in both the Regge propagators because the $\rho$ and
$\pi$ mesons are degenerate in pion photoproduction within
a Regge model~\cite{Guidal:1997hy}. Though a phase factor,  
$\exp(-i\pi\alpha_{\rho(\pi)}(t))$, can also be included, we
find that the results change only slightly.

The Regge amplitude $T_R$ can be expressed in terms of the 
individual invariant amplitudes combined with the Regge propagators
\begin{align}
T_R = \mathcal M_\rho \cdot (t-M_\rho^2) \cdot R_\rho(t) \cdot C_\rho(t) + 
\mathcal M_\pi \cdot (t-M_\pi^2) \cdot R_\pi(t) \cdot C_\pi(t),
\label{eq:TotRegAmpl}
\end{align}
where $C_\rho (t)$ and $C_\pi(t)$ are called scale
factors~\cite{Kim:2015ita,Titov:2008yf}, which are employed to fit
the experimental data based on the $\pi^- p \to \phi n$ reaction 
at high energies.

Now, we consider the resonance contribution from the pentaquark states
$P_c^0(4380)$ and $P_c^0(4350)$. 
The exact quantum numbers of these two pentaquark
states are not known. However, a previous study~\cite{Aaij:2015tga}
suggested that the spins and parities for $P_c(4380)$ and $P_c(4450)$ should be
$(3/2^-,5/2^+)$, respectively, for the best fit in partial-wave 
analysis. In addition, the combinations of 
$(3/2^+,5/2^-)$ and $(5/2^+,3/2^-)$ provide acceptable solutions
for the pentaquark states. 
We also consider the case of $(5/2^-,3/2^+)$ in 
our calculation.

The effective Lagrangians for the $P_c N \pi$ vertex shown in 
Fig.~\ref{FIG2}(b) are given as~\cite{Oh:2011}
\begin{align}
\mathcal L_{P_c N \pi}^{3/2^\pm}  &=
\frac{g_{P_c N \pi}}{M_\pi} \bar N \Gamma^{(\mp)} \bm{\tau} \cdot 
\partial_\mu \bm{\pi} P_c^\mu + \mathrm{H.c.},                         \cr
\mathcal L_{P_c N \pi}^{5/2^\pm} &= 
i \frac{g_{P_c N \pi}}{M_\pi^2} \bar N \Gamma^{(\pm)} \bm{\tau} \cdot 
\partial_\mu \partial_\nu \bm{\pi} P_c^{\mu\nu} + \mathrm{H.c.},
\label{eq:ResLag1}
\end{align}
where we ignore the off-shell
part of the Rarita-Schwinger fields because the resonances are almost
on mass shell. The following notations are used 
\begin{align}
\Gamma^{(\pm)} = \left(
\begin{array}{c} 
\gamma_5 \\ \mathbf{1}
\end{array} \right),
\qquad
\Gamma_\mu^{(\pm)} = \left(
\begin{array}{c}
\gamma_\mu \gamma_5 \\ \gamma_\mu 
\end{array} \right).
\end{align}
After we obtain the branching ratio of $P_c \to N\pi$, we can easily
determine the coupling constants $g_{P_c N \pi}$ from the decay
widths~\cite{Oh:2011} 
\begin{align}
\Gamma (P_c^{3/2^\pm} \to N \pi) &= \frac{g_{P_c N \pi}^2}{4 \pi} 
\frac{p_N^3}{M_\pi^2 M_{P_c}} (E_N \pm M_N),                           \cr  
\Gamma (P_c^{5/2^\pm} \to N \pi) &= \frac{2}{5} \frac{g_{P_c N \pi}^2}{4 \pi} 
\frac{p_N^5}{M_\pi^4 M_{P_c}} (E_N \mp M_N),
\label{eq:DW1}
\end{align}
where $E_N = (M_{P_c}^2 + M_N^2 - M_\pi^2) / (2M_{P_c})$ and
$p_N = \sqrt{E_N^2 -M_N^2}$. Unfortunately, the branching ratios are
not known at present, so we have to rely on previous experimental
information on the upper limit on the total cross section for 
$\pi N\to J/\psi N$~\cite{Jenkins:1977xb,Chiang:1986gn}.

The effective Lagrangians for the $P_c N J/\psi$ vertex can be expressed
as~\cite{Kim:2011rm}
\begin{align}
\mathcal{L}_{P_c N \psi}^{3/2^\pm}                                    
&= i\overline {P_c}_\mu \left[ 
\frac{g_1}{2M_N} \Gamma_\nu^{(\pm)} N \mp 
\frac{ig_2}{(2M_N)^2} \Gamma^{(\pm)} \partial_\nu N \pm 
\frac{ig_3}{(2M_N)^2} \Gamma^{(\pm)} N \partial_\nu 
\right] \psi^{\mu\nu} + \mathrm{H.c.},                                \cr  
\mathcal{L}_{P_c N \psi}^{5/2^\pm}                                           
&= \overline {P_c}_{\mu\alpha} \left[   
\frac{g_1}{(2M_N)^2} \Gamma_\nu^{(\mp)} N \pm 
\frac{ig_2}{(2M_N)^3} \Gamma^{(\mp)} \partial_\nu N \mp 
\frac{ig_3}{(2M_N)^3} \Gamma^{(\mp)} N \partial_\nu 
\right] \partial^\alpha \psi^{\mu\nu} \cr
&\;\;\;\;\;\; + \mathrm{H.c.}.
\label{eq:ResLag2}
\end{align}
The heavy pentaquark states play a dominant role near the
threshold region, so we only consider the first term in
Eq.~(\ref{eq:ResLag2}). Thus, the decay widths for the heavy
pentaquarks are written as~\cite{Oh:2011} 
\begin{align}
\Gamma (P_c^{3/2^\pm} \to N J/\psi) =& 
\frac{g_{P_c N J/\psi}^2}{12\pi} \frac{p_N}{M_{P_c}} (E_N \mp M_N)         \cr   
&\times
[2E_N(E_N \pm M_N) + (M_{P_c} \pm M_N)^2 + 2M_{J/\psi}^2],               \cr
\Gamma (P_c^{5/2^\pm} \to N J/\psi) =& 
\frac{g_{P_c N J/\psi}^2}{60\pi} \frac{p_N^3}{M_{P_c}} (E_N \pm M_N)  \cr   
&\times 
[4E_N(E_N \mp M_N) + (M_{P_c} \mp M_N)^2 + 4M_{J/\psi}^2], 
\label{eq:DW2}
\end{align}
which we can use to determine the coupling constants for the $P_c$s. 
The kinematic variables $E_N$ and $p_N$ in Eq.~(\ref{eq:DW2}) are
defined as $E_N = (M_{P_c}^2 + M_N^2 - M_{J/\psi}^2) / (2M_{P_c})$
and $p_N = \sqrt{E_N^2 -M_N^2}$.

Finally, we have the following expressions for the $s$ channel  
\begin{align}
\mathcal{M}_{P_c(3/2^+)}^\mu 
=& I_{P_c} \frac{ig_{P_c N J/\psi}}{2M_N} \frac{g_{P_c N \pi}}{M_\pi} 
\frac{1}{s-M_{P_c}^2} \gamma_5 \gamma_\nu  
(k_2^\alpha g^{\mu\nu} - k_2^\nu g^{\mu\alpha})
\Delta_{\alpha\beta}(P_c,k_1 +p_1) k_1^\beta,                              \cr
\mathcal{M}_{P_c(3/2^-)}^\mu 
=& I_{P_c} \frac{ig_{P_c N J/\psi}}{2M_N} \frac{g_{P_c N \pi}}{M_\pi} 
\frac{1}{s-M_{P_c}^2} \gamma_\nu  
(k_2^\alpha g^{\mu\nu} - k_2^\nu g^{\mu\alpha}) 
\Delta_{\alpha\beta}(P_c,k_1 +p_1) k_1^\beta \gamma_5,                     \cr   
\mathcal{M}_{P_c(5/2^+)}^\mu 
=& - I_{P_c} \frac{ig_{P_c N J/\psi}}{(2M_N)^2} \frac{g_{P_c N \pi}}{M_\pi^2} 
\frac{1}{s-M_{P_c}^2} \gamma_\nu k_2^{\alpha_2}
(k_2^{\alpha_1} g^{\mu\nu} - k_2^\nu g^{\alpha_1 \mu})                          \cr  
 &\times 
\Delta_{\alpha_1 \alpha_2 ; \beta_1 \beta_2}(P_c,k_1 +p_1) 
k_1^{\beta_1}  k_1^{\beta_2} \gamma_5,                                     \cr
\mathcal{M}_{P_c(5/2^-)}^\mu
=& - I_{P_c} \frac{ig_{P_c N J/\psi}}{(2M_N)^2} \frac{g_{P_c N \pi}}{M_\pi^2} 
\frac{1}{s-M_{P_c}^2} \gamma_5 \gamma_\nu k_2^{\alpha_2}
(k_2^{\alpha_1} g^{\mu\nu} - k_2^\nu g^{\alpha_1 \mu})                         \cr 
 &\times 
\Delta_{\alpha_1 \alpha_2 ; \beta_1 \beta_2}(P_c,k_1 +p_1)
  k_1^{\beta_1}  k_1^{\beta_2},
\label{eq:InvAmp2}   
\end{align}
where the different spins and parities of the $P_c$ states are assumed.
The isospin factors are given by $I_{P_c} = \sqrt{2}$.
Given the decay widths of the $P_c$ states, the propagators of
the pentaquark states should be modified to $M_{P_c} \to (M_{P_c} -
i\Gamma_{P_c}/2$). Previous studies~\cite{Kim:2011rm,Kim:2012pz} provided
the explicit expressions for $\Delta_{\alpha\beta}$ and
$\Delta_{\alpha_1,\alpha_2;\beta_1\beta_2}$ in Eq.~(\ref{eq:InvAmp2}).
The relevant hadrons are spatially extended, so we consider the 
phenomenological form factors in the $s$ channel 
\begin{align}
F_{P_c}(s) = \left( \frac{\Lambda^4}{\Lambda^4 + (s-M_{P_c}^2)^2} \right)^2,
\label{eq:FF}
\end{align}
where the cutoff masses are selected as $\Lambda=1.0$ GeV.
The cutoff masses actually play no crucial roles in this
calculation because the pentaquark states lie almost near the threshold
region.
\section{Results}
\label{SecIII}
First, we study the background contribution ($\rho$ and $\pi$ Reggeon
exchanges) to the total cross sections for both the $\pi^- p \to \phi
n$ and $\pi N\to J/\psi N$ reactions.
\begin{figure}[thp]
\includegraphics[width=10cm]{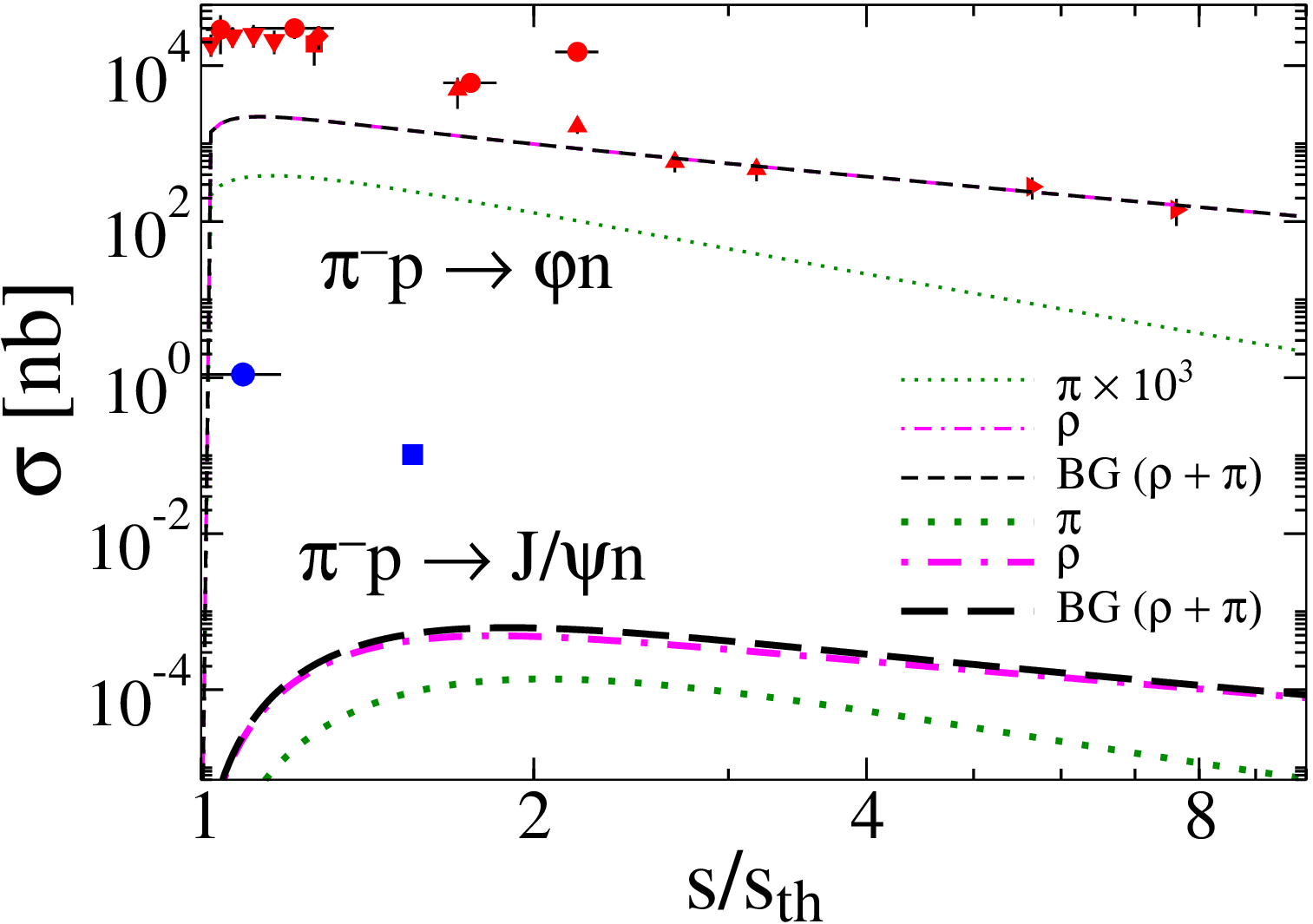}
\caption{(color online) Background (BG) contributions to the total 
cross sections for $\pi^- p \to (\phi,J/\psi)n$ with $\rho$ and $\pi$
Reggeon exchanges.
The experimental data on $\pi^- p \to \phi n$ are taken from
Dahl et al.~\cite{Dahl:1967pg} (red circles, squares, and triangles)
and the data on the upper limit of the $\pi^- p \to J/\psi n$ reaction
are taken from Jenkins et al.~\cite{Jenkins:1977xb} (blue circle) and
Chiang et al.~\cite{Chiang:1986gn} (blue square).}   
\label{FIG3}
\end{figure}
Figure~\ref{FIG3} shows the effects of $\rho$ and $\pi$ Reggeon exchanges 
on both the reactions. Note that the total cross sections are drawn as
a function of $s/s_{\mathrm{th}}$ so we can easily compare the
total cross section of the $\pi^- p \to \phi n$ reaction with that of
$J/\psi$ production. $s_{\mathrm{th}}$ denotes the threshold
value of $s$, i.e., 
$s_{\mathrm{th}}^s=(M_\phi + M_n)^2=3.84\,\mathrm{GeV}^2$ and
$s_{\mathrm{th}}^c=(M_{J/\psi} + M_n)^2=16.3\,\mathrm{GeV}^2$ for the
$\pi^- p \to \phi n$ and $\pi^- p \to J/\psi n$ reactions, respectively. 
The $\rho$ Reggeon exchange dominates the
$\pi$ Reggeon exchange, which is expected due to the relatively smaller
values of $g_{V\pi\pi}$ compared with those of
$g_{V\rho\pi}$, where $V$ generically represents $\phi$ and $J/\psi$ 
mesons (see Table~\ref{TABLE1}). The Regge approach is known 
to describe the experimental data well at higher energies, so we fit the
scale factor $C_{\rho(\pi)}(t)$ defined in Eq. (\ref{eq:TotRegAmpl}) 
such that it explains the total cross section for $\pi^-
p\to \phi n$~\cite{Dahl:1967pg} in the higher energy region.
We use the form of $C_{\rho(\pi)}(t) = 0.5/(1-t/\Lambda^2)^2$ with the
cutoff mass $\Lambda$ = 1 GeV fixed in Eq.~(\ref{eq:FF}) to avoid
additional ambiguity. We use the same form of the scale factor to
obtain the total cross section for the $\pi^- p \to J/\psi n$
reaction. The results of the background contribution lie below the
experimental upper limit~\cite{Jenkins:1977xb, Chiang:1986gn} on the
total cross section of the $\pi^- p \to J/\psi n$ reaction.  

As shown in Fig.~\ref{FIG3}, we find that the magnitude of the
background contribution to the total cross section for $\pi^- p
\to J/\psi n$ is about $10^6$ times smaller than that for $\pi^- p
\to \phi n$. 
This is due mostly to the greatly suppressed value of 
$g_{J/\psi\rho\pi}/M_{J/\psi}$ in Eq.~(\ref{eq:Lag1}) compared with 
$g_{\phi\rho\pi}/M_\phi$ in the case of dominant $\rho$-meson Reggeon 
exchange: $( g_{\phi\rho\pi}/M_\phi )^2 \simeq 
2.5 \cdot 10^5 \times (g_{J/\psi\rho\pi}/M_{J/\psi})^2$.
It is interesting to compare the current results with those of 
previous theoretical studies. For example, Kodaira and
Sasaki~\cite{Kodaira:1979sf} estimated the total cross section for the
$\pi^- p \to J/\psi n$ reaction by using generalized Veneziano models
many years ago. The results were obtained as $\sigma(\pi^- p
\to J/\psi n)= (1.1,\,0.44)\,\mathrm{pb}$ at $p_{\mathrm{lab}} = (50,\,
100)\, \mathrm{GeV}/c$, respectively, and we  
obtained $(0.17,\, 0.071)\, \mathrm{pb}$ at the corresponding
momenta. Thus, the orders of magnitude appear to be similar to each
other. However, Wu and Lee~\cite{Wu:2013xma} predicted about 1.5 nb
near the threshold region (W $\sim$ 4.2 GeV) within a coupled-channel
model. Lu et al.~\cite{Lu:2015fva} computed the total cross section
for the $\pi^- p \to J/\psi n$ reaction by considering the heavy pentaquark
states, where they determined the background contribution to the
total cross section in the vicinity of the $P_c$ resonances as in the order of
$10-100$ nb, which is about $10^4-10^5$ times larger than those obtained in the
present study.

In the case of the $\pi N\to \phi N$ reaction, the contributions of
the nucleon resonances have roles at low energies, as
studied by Xie et al.~\cite{Xie:2007qt}, who focused on the role of $N^*(1535)$.
Similarly, the heavy pentaquark resonances will make 
a specific contribution to the $\pi^- p \to J/\psi n$ reaction near the 
threshold.  
\begin{figure}[hp]
\vspace{2em}
\includegraphics[width=10.00cm]{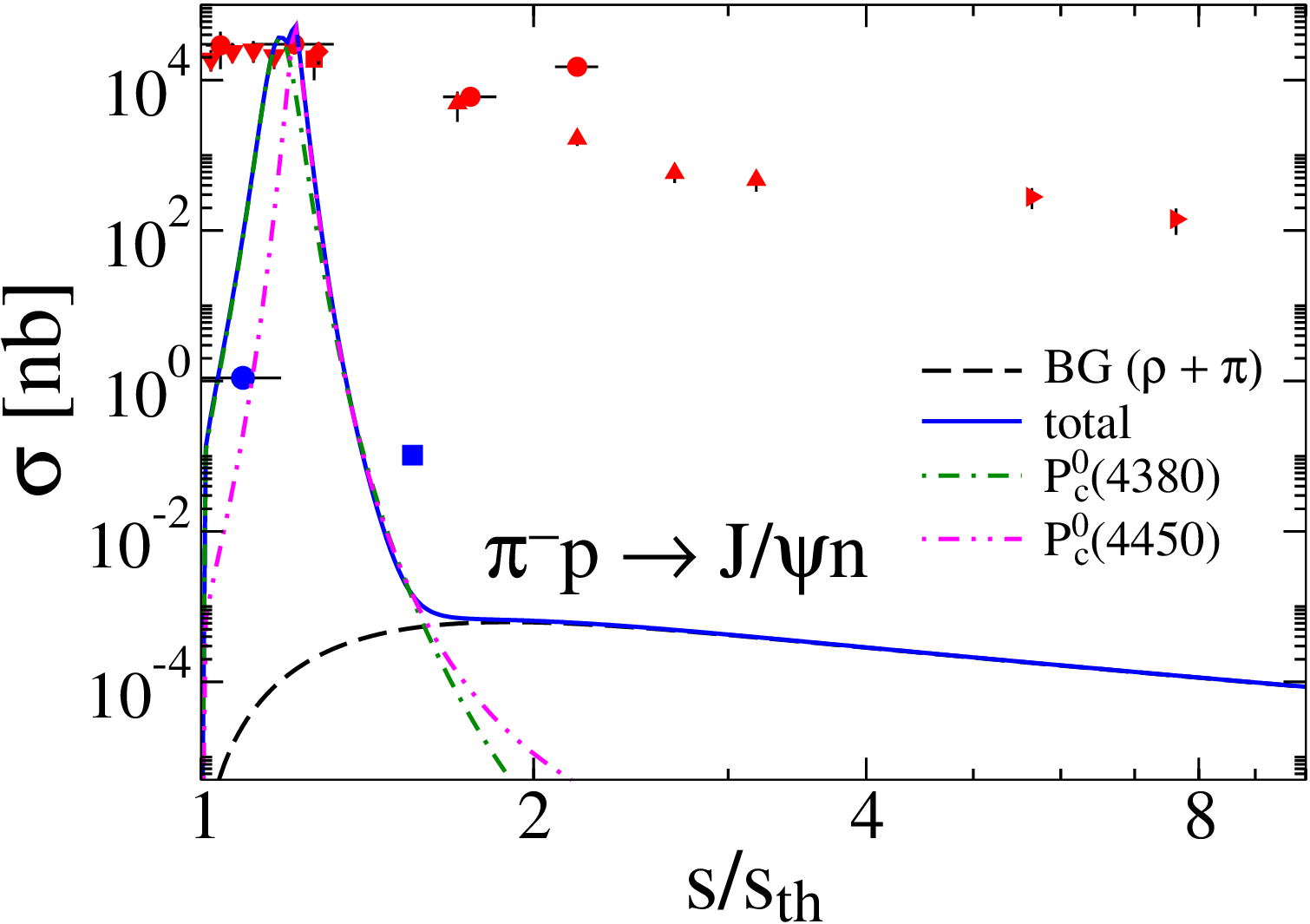}
\caption{(color online) Total cross section for 
$\pi^- p \to J/\psi n$ under the assumption that
$\mathcal{B}(P_c^0 \to J/\Psi N) = 40$\% and $\mathcal{B}(P_c^0 \to
\pi N) = 6$\% extracted from Wu et al.~\cite{Wu:2010jy}. 
The spin and parity are given as $J^P = (3/2^-,5/2^+)$ for 
($P_c^0(4380)$,\,$P_c^0(4450)$), respectively.
The notations for the experimental data are the same as those given in
Fig.~\ref{FIG3}.}  
\label{FIG4}
\end{figure}
When considering the heavy pentaquark resonances, 
we encounter a problem with determining the coupling constants
due to the unknown branching ratios of the $P_c$ states relative to any 
other channels. Wu et al.~\cite{Wu:2010jy} predicted new resonances
with the $c\bar c$ component based on a unitarized coupled-channel
formalism before observations of the $P_c$ states were
reported. The branching ratios of the predicted states for the
$J/\psi N$ and $\pi N$ decay channels were proposed as about
$40~\%$ and $6~\%$, respectively. Assuming that they are the heavy 
pentaquark states announced recently, we find that the magnitude of 
the total cross section reaches about $10^4$ nb, as depicted in
Fig.~\ref{FIG4}, which is even closer to the magnitude of the total
cross section for the $\pi^- p \to \phi n$ reaction. Moreover, it
exceeds the experimental upper limit on the $\pi^- p \to J/\psi n$
reaction~\cite{Jenkins:1977xb,Chiang:1986gn} by approximately $10^4$
times. Note that we have set the quantum numbers of the $P_c$
states as $J^P = (3/2^-,5/2^+)$, which are mostly favorable.
We consider different combinations of the spin and
parity for the $P_c$ states, but we cannot avoid this large result for the
total cross section.    

We may follow another suggestion proposed 
by Wang et al.~\cite{Wang:2015jsa}, where $\mathcal{B}(P_c^+ \to J/\Psi p) =
5~\%$ was assumed for the study of $J/\psi$ photoproduction. As a
result, Wang et al.~\cite{Wang:2015jsa} were able to describe the old 
experimental data even when the heavy pentaquark states were
considered. Following the same line of reasoning, we consider the
experimental upper limit for the $\pi N\to J/\psi N$ reaction, which
is given as around 1 nb~\cite{Jenkins:1977xb}. This implies that the
branching ratio of $P_c\to \pi N$ should be very small:
$\mathcal{B}(P_c^0 \to \pi N) \sim 10^{-5}$. Thus, we can obtain the 
corresponding coupling constants using Eqs.~(\ref{eq:DW1}) and
(\ref{eq:DW2}). The results are listed in Table~\ref{TABLE2}.
\begin{table}[thp]
\begin{tabular}{ccc|cccc} \hline\hline
&$\mathrm{Coupling}$
&$\hspace{0.2cm}\mathrm{State}\hspace{0.2cm}$
&$\hspace{0.6cm}3/2^+\hspace{0.6cm}$
&$\hspace{0.6cm}3/2^-\hspace{0.6cm}$
&$\hspace{0.6cm}5/2^+\hspace{0.6cm}$
&$\hspace{0.6cm}5/2^-\hspace{0.6cm}$ \\
\hline
&$g_{P_c N \pi}$&$P_c(4380)$
&$2.74 \cdot 10^{-4}$&$4.23 \cdot 10^{-4}$
&$4.47 \cdot 10^{-5}$&$2.89 \cdot 10^{-5}$ \\
&&$P_c(4450)$ 
&$1.17 \cdot 10^{-4}$&$1.79 \cdot 10^{-4}$
&$1.86 \cdot 10^{-5}$&$1.21 \cdot 10^{-5}$ \\ 
&$g_{P_c N J/\psi}$&$P_c(4380)$
&0.771&0.345&1.53&3.61 \\
&&$P_c(4450)$
&0.291&0.141&0.568&1.24 \\
\hline\hline
\end{tabular}
\caption{Coupling constants for the heavy pentaquark states with 
each $J^P$ assignment. The branching ratios for the pentaquarks are
assumed to be $\mathcal{B}(P_c^0 \to \pi N) = 10^{-5}$  
and $\mathcal{B}(P_c^0 \to J/\Psi N) = 0.05$.} 
\label{TABLE2}
\end{table}

\begin{figure}[thp]
\vspace{1em}
\begin{tabular}{cc}
\includegraphics[width=7.00cm]{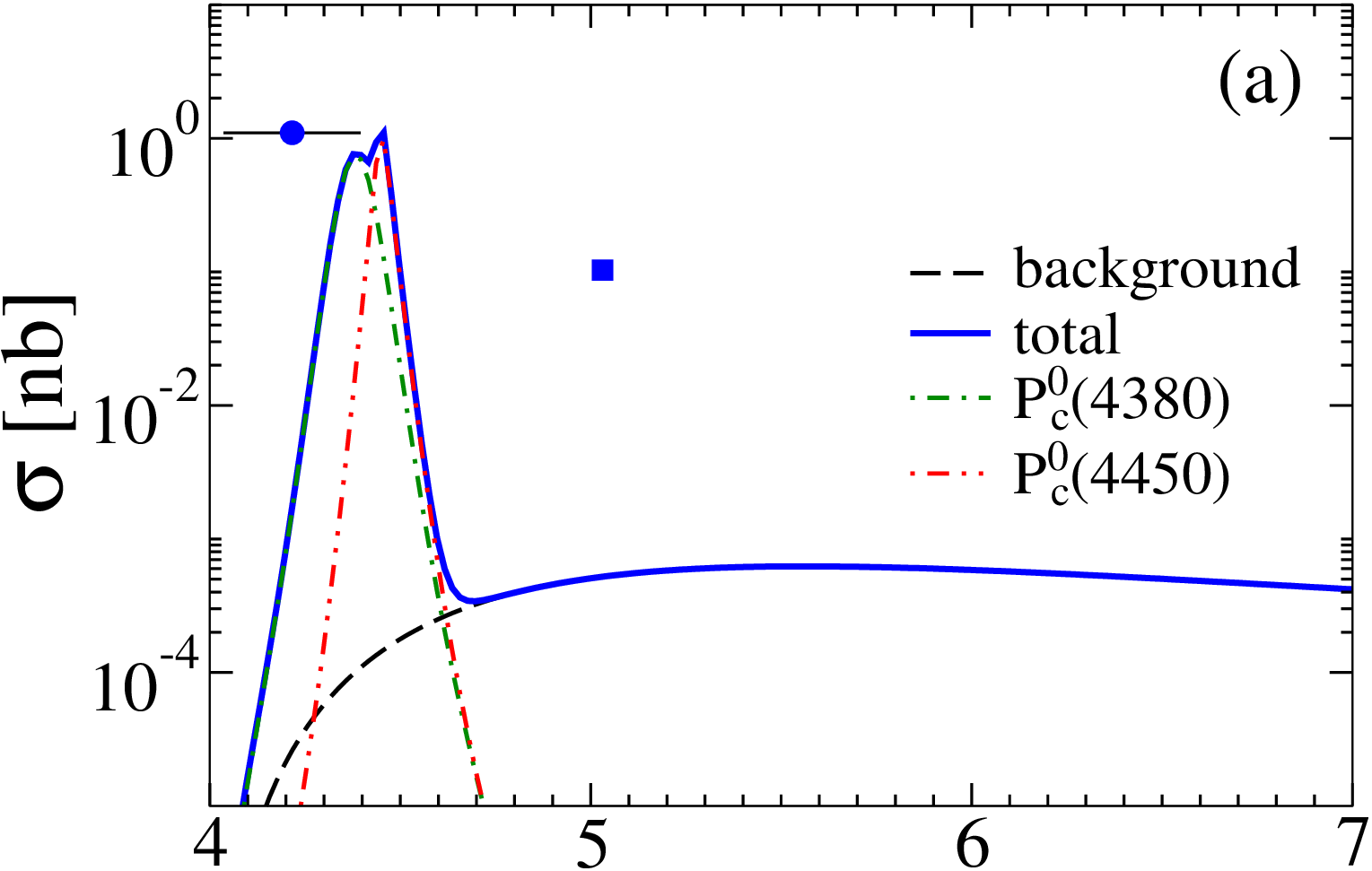}\,\,\,\,\,
\includegraphics[width=7.00cm]{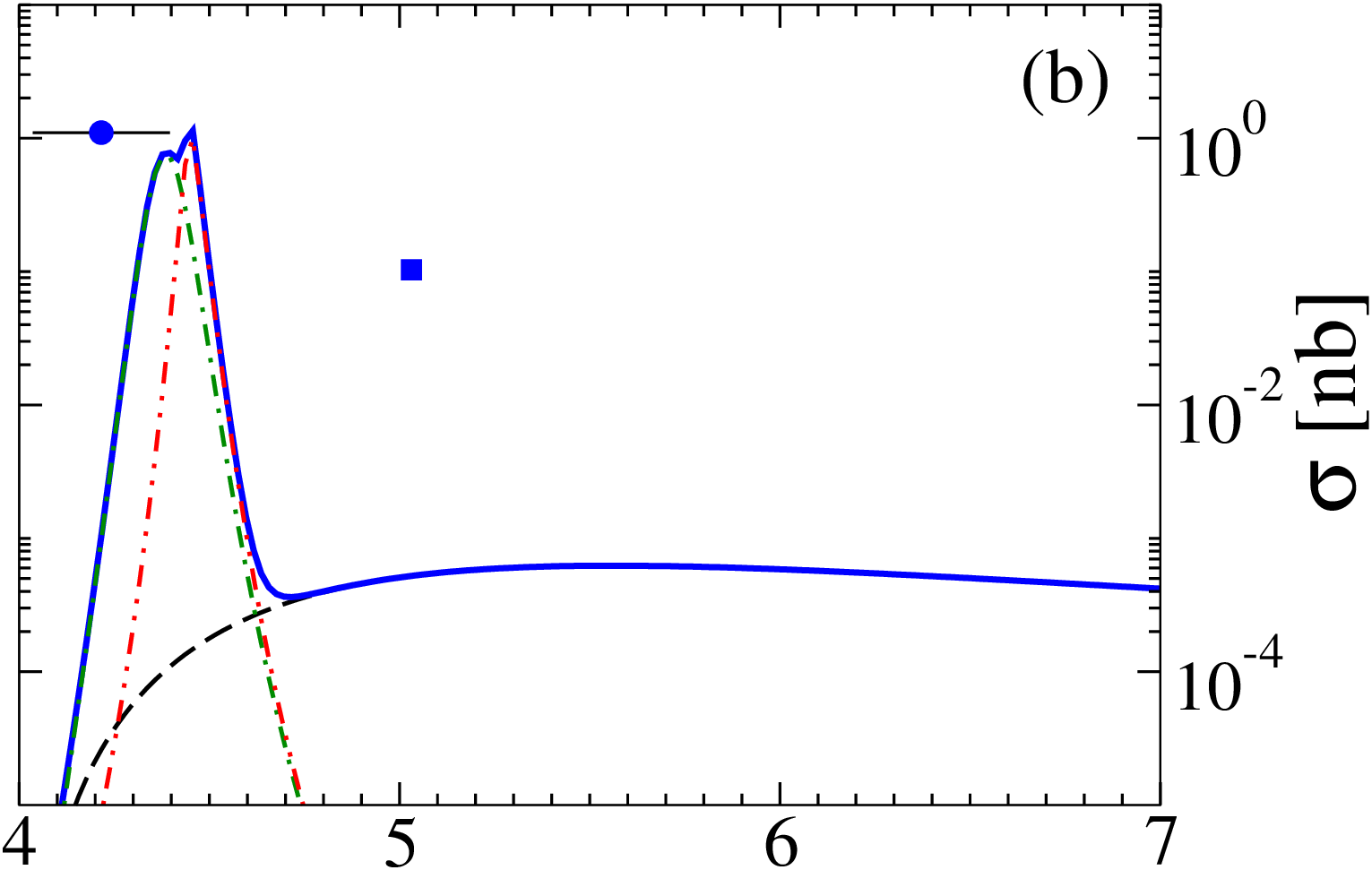}\vspace{1.75em} \\
\includegraphics[width=7.00cm]{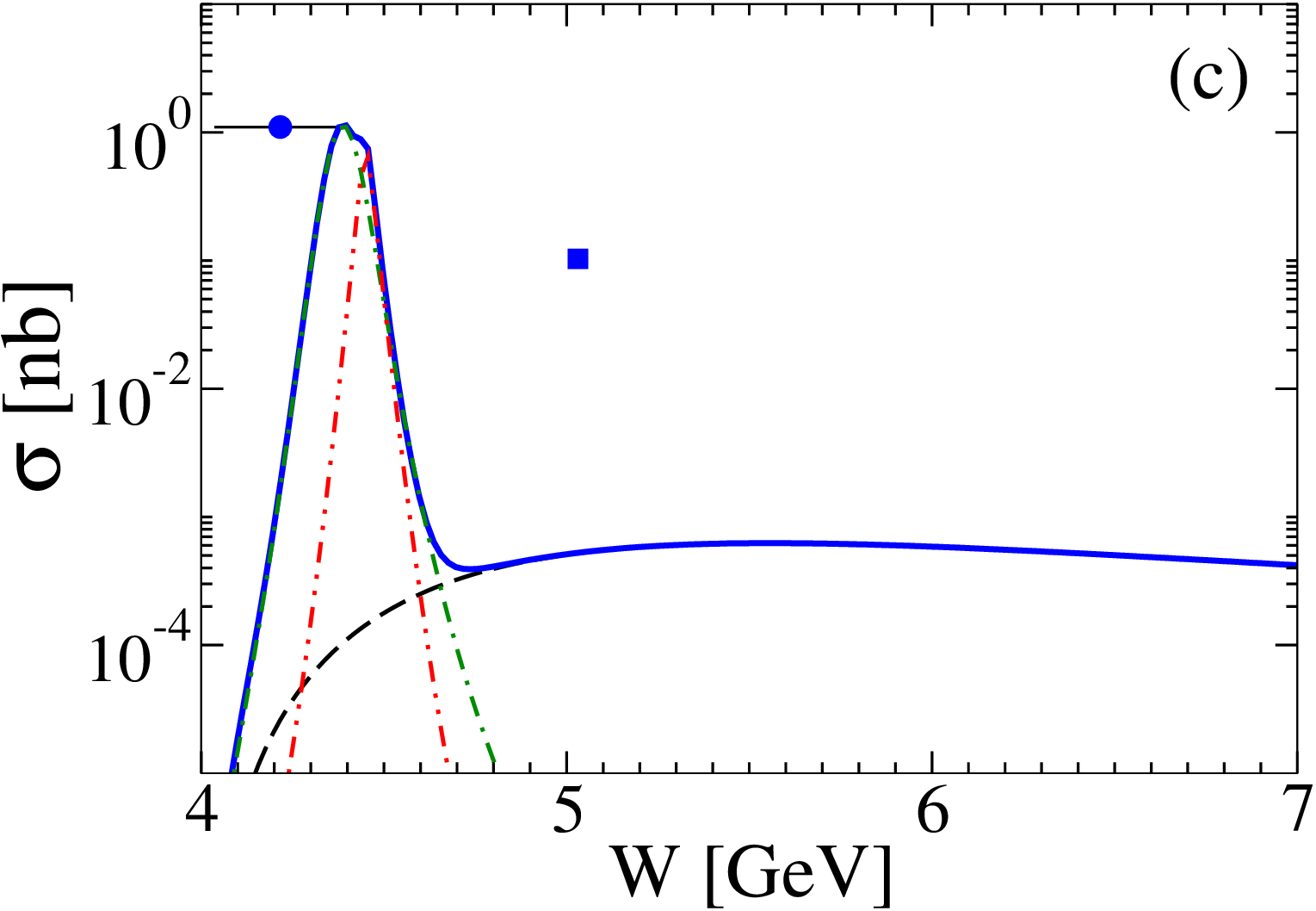}\,\,\,\,\,
\includegraphics[width=7.00cm]{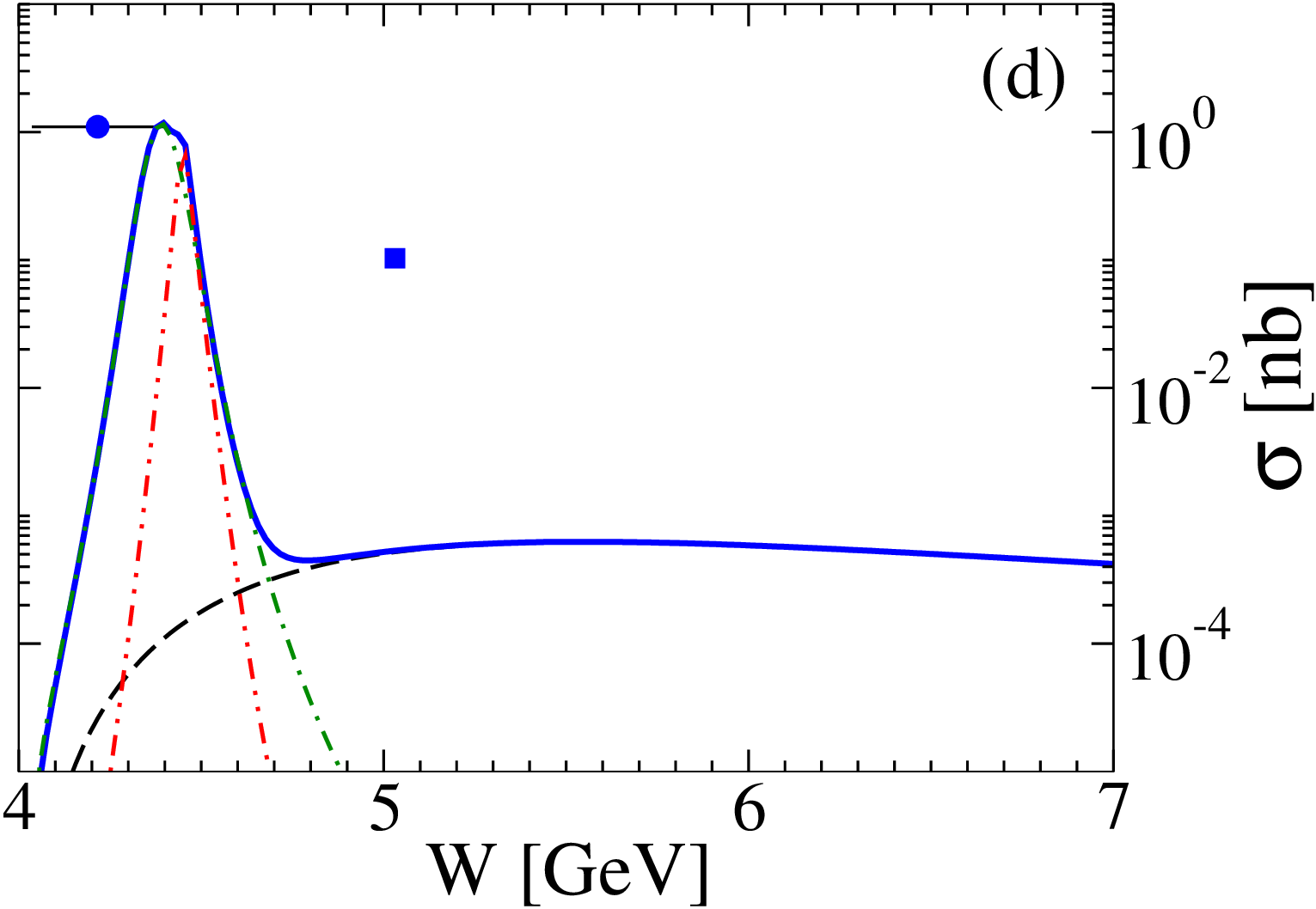}
\end{tabular}
\caption{Total cross sections for $\pi^- p \to J/\psi n$.
The spin-parity selections of (a--d) are $J^P = (3/2^-,5/2^+), 
(3/2^+,5/2^-), (5/2^+,3/2^-), (5/2^-,3/2^+)$ for 
($P_c^0(4380)$,\,$P_c^0(4450)$), respectively.
The experimental data come from Jenkins et al.~\cite{Jenkins:1977xb} 
(blue circle) and Chiang et al.~\cite{Chiang:1986gn} (blue square).}
\label{FIG5}
\end{figure}

Figure~\ref{FIG5} shows our results for the total cross
section as a function of the center of mass (CM) energy $W$, where
different combinations of the spin and parity are considered for the
heavy pentaquark states. As illustrated in Fig.~\ref{FIG5}, the
results are not sensitive to the selection of the spin and parity 
for the heavy pentaquark states. The peaks of the $P_c$ states reach
the experimental upper limit, i.e., about 1 nb.  
$P_c(4380)$ has a broad width
($\Gamma_{P_c(4380)}\approx 210\,\mathrm{MeV}$), so its peak
overlaps with that of $P_c(4450)$, where the width is
$\Gamma_{P_c(4450)}\approx 40\,\mathrm{MeV}$. Thus, it may be very
difficult to directly distinguish $P_c(4380)$ from 
$P_c(4450)$ based on the $\pi^- p \to J/\psi n$ reaction.

We note that a larger value of the
branching ratio could be considered for the $P_c\to J/\Psi$ decay, where we suppress that
of $P_c^0 \to \pi N$ to examine the dependence of the total cross
section on them. For example, we can choose
$\mathcal{B}(P_c^0 \to J/\Psi N) = 0.5$, which is 10 times larger
than the value used in the present study. By contrast,
we may take $\mathcal{B}(P_c^0 \to \pi N) = 10^{-6}$, which is 10 times
smaller than the value in the present study. However, we obtain almost the
same numerical results.

\begin{figure}[htp]
\vspace{2em}
\includegraphics[width=10.00cm]{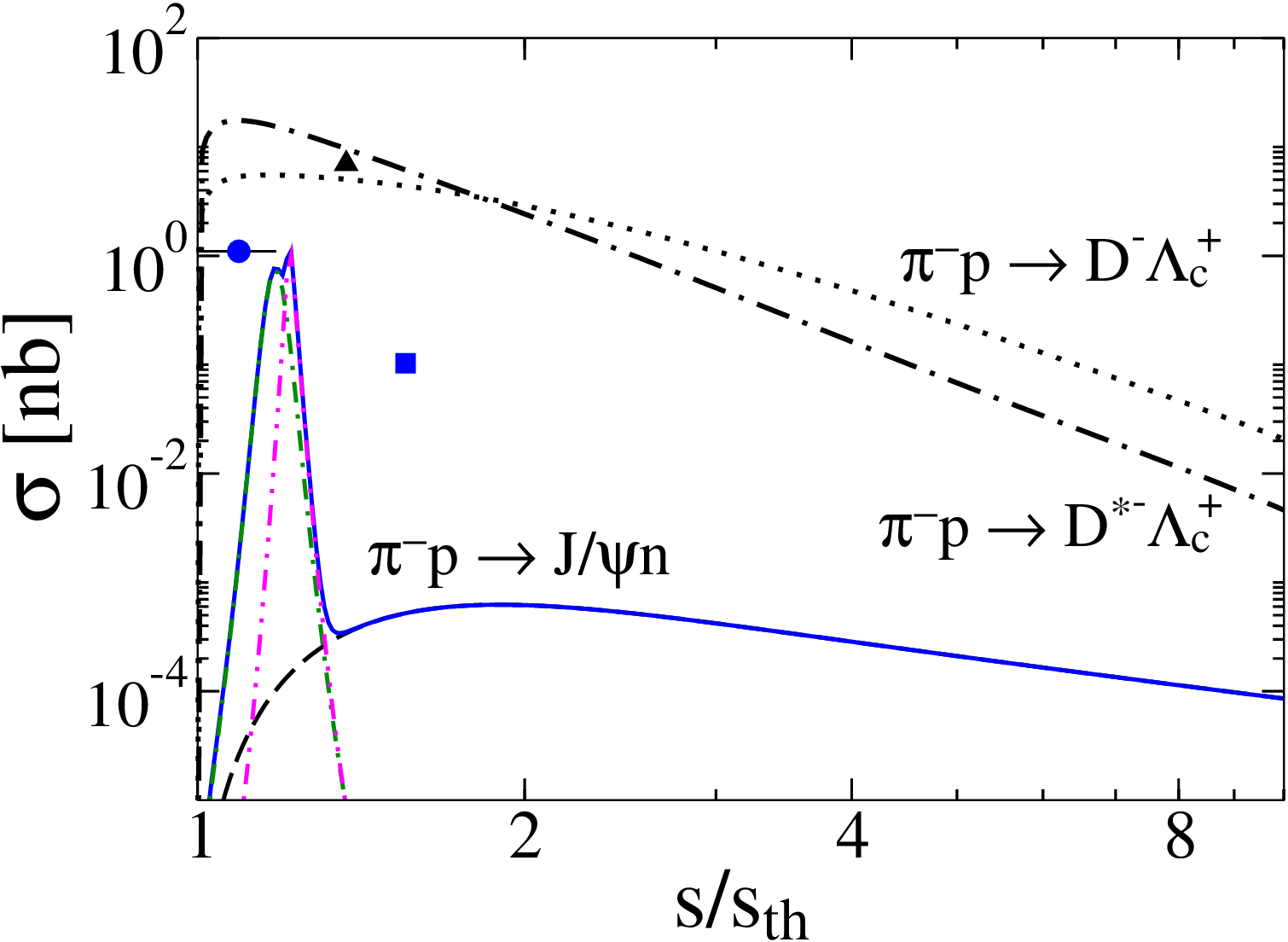}
\caption{(color online) Total cross section for 
$\pi^- p \to J/\psi n$. The notation is the same as that employed in
Fig.~\ref{FIG5}.  The spin and parity are given as $J^P =
(3/2^-,5/2^+)$ for ($P_c^0(4380)$,\,$P_c^0(4450)$), respectively. 
The experimental data on the upper limit come from 
Jenkins et al.~\cite{Jenkins:1977xb} (blue circle) and Chiang et
al.~\cite{Chiang:1986gn}  
(blue square). The black triangle represents the
upper limit on $D^{*-}\Lambda_c^+$
production~\cite{Christenson:1985ms}. The dotted curve depicts the
results of the total cross section for $\pi^-p\to D^-\Lambda_c^+$,
and the dot-dashed curve illustrates that for $\pi^-p\to 
D^{*-}\Lambda_c^+$. The contribution of the heavy pentaquark is not
included in the results for these two reactions.}
\label{FIG6}
\end{figure}

In addition to the hidden-charm production process investigated in
this study, it is also very interesting to investigate the production of
the heavy pentaquarks in the open-charm channel. 
Previously, we performed studies based on 
the $\pi^- p \to D^{*-}\Lambda_c^+$~\cite{Kim:2015ita} and 
$\pi^- p \to D^-\Lambda_c^+$~\cite{Kim:2016abc} reactions without 
considering the pentaquark states. We compare the corresponding
results for the total cross sections with those obtained in the
present study in Fig.~\ref{FIG6}. As expected, due to OZI 
suppression, the total cross section for $J/\Psi n$ production
is approximately $10^2-10^4$ smaller than those for 
$D^{*-}\Lambda_c^+$ and $D^-\Lambda_c^+$ production, excluding the
resonance region. This indicates that if the pentaquark states are
strongly coupled to the $D^{*-}\Lambda_c^+$ and/or $D^-\Lambda_c^+$
channels, it might be easier to find evidence for the existence of the 
heavy pentaquarks in open-charm processes. 

Assuming that $\rho$-meson exchange is dominant, we find 
that the other isospin channels for the $\pi N\to J/\psi N$ reactions are
related to each other by the isospin factors given as  follows
\begin{align}
\sigma (\pi^- p \to J/\psi n) = \sigma (\pi^+ n \to J/\psi p) = 
2 \,\sigma (\pi^0 n \to J/\psi n) = 2 \,\sigma (\pi^0 p \to J/\psi p). 
\end{align}
However, considering that $J/\psi$ cannot decay into two
neutral pions and charged mesons are not allowed to be exchanged, then the
mechanism of the $\pi^0 N\to J/\psi N$ reactions should differ
from that of the $\pi^\pm N\to J/\psi N$ processes. Note that the $\pi^0$
beam is not suitable for use in the experimental production of
any hadrons because of its neutral nature and short lifetime. 
However, the $\pi^+ n \to J/\psi p$ reaction provides an
opportunity to study the existence of the charged heavy pentaquark
$P_c^+$, and the present study considers the neutral $P_c^0$.

\begin{figure}[thp]
\vspace{1.1em}
\begin{tabular}{cc}
\includegraphics[width=7.00cm]{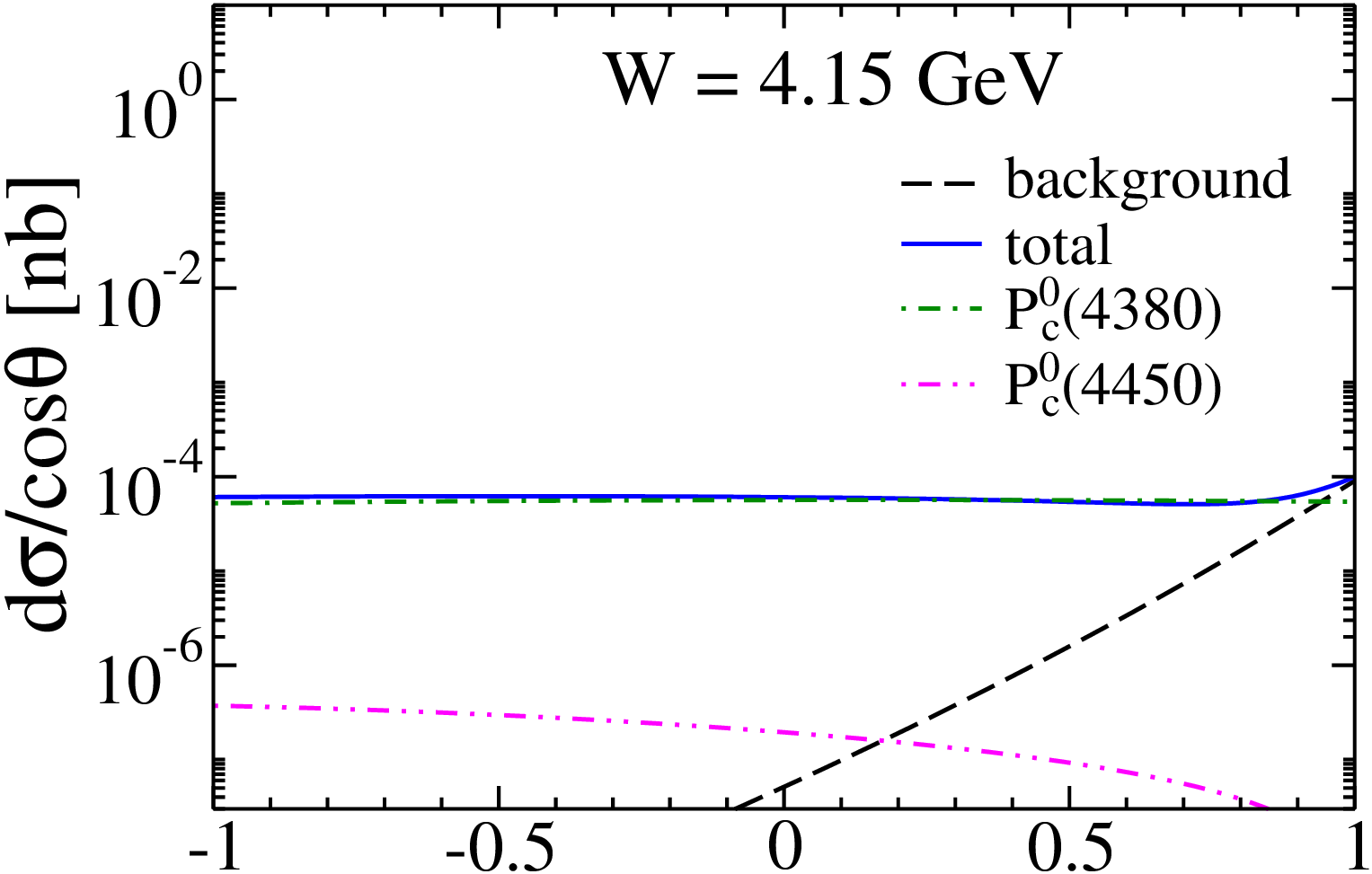}\,\,\,\,\,
\includegraphics[width=7.00cm]{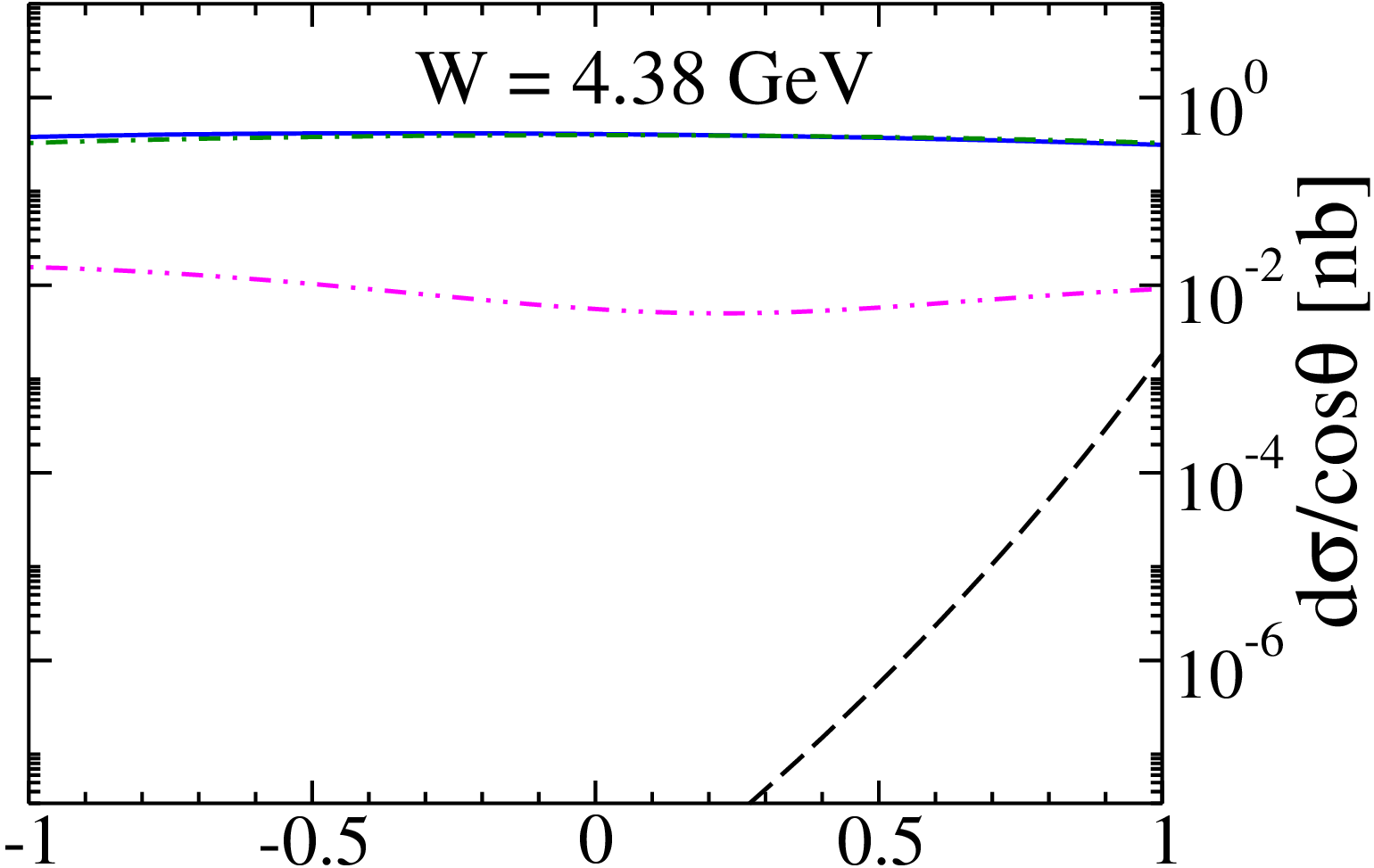}\vspace{1.75em} \\
\includegraphics[width=7.00cm]{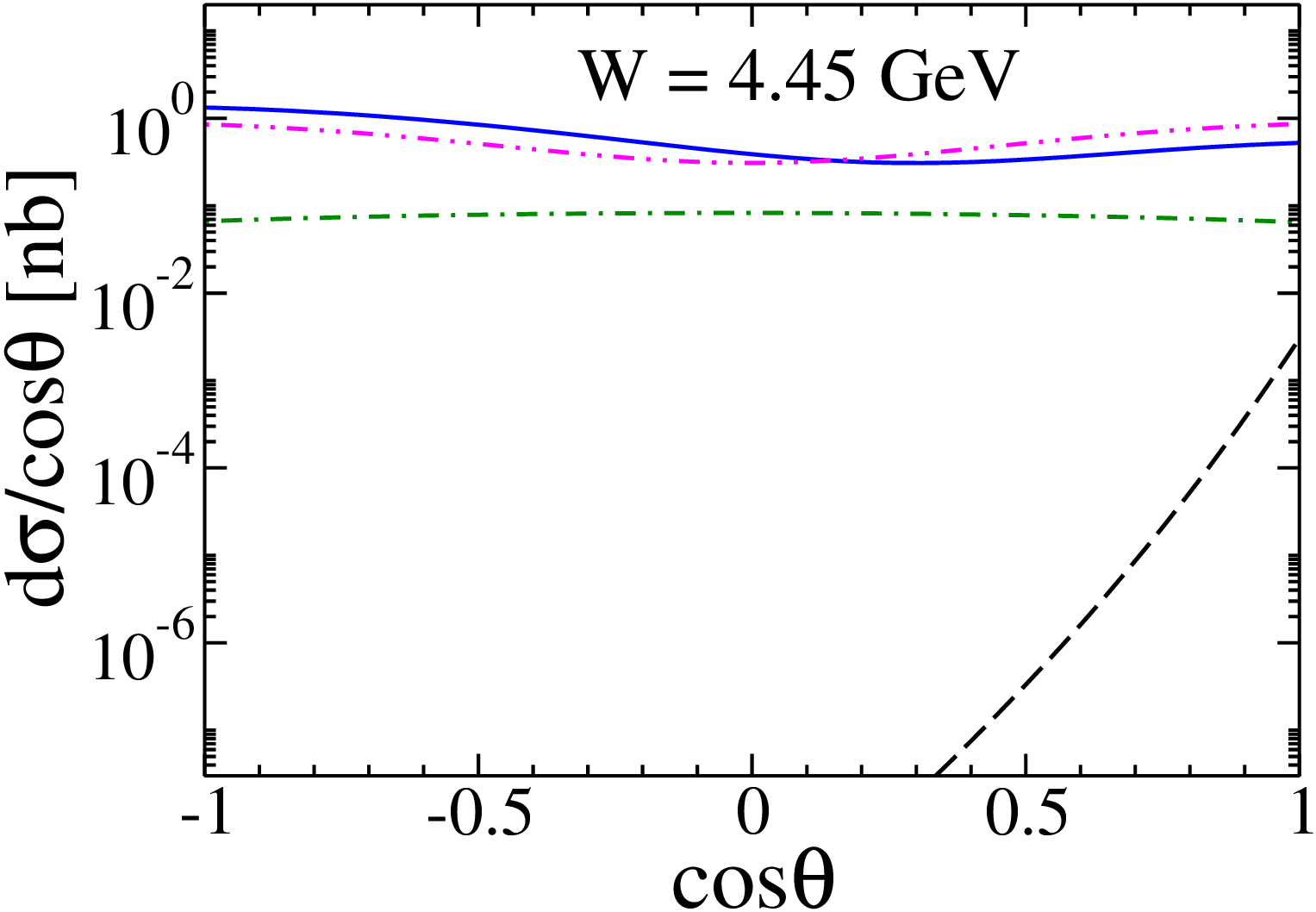}\,\,\,\,\,
\includegraphics[width=7.00cm]{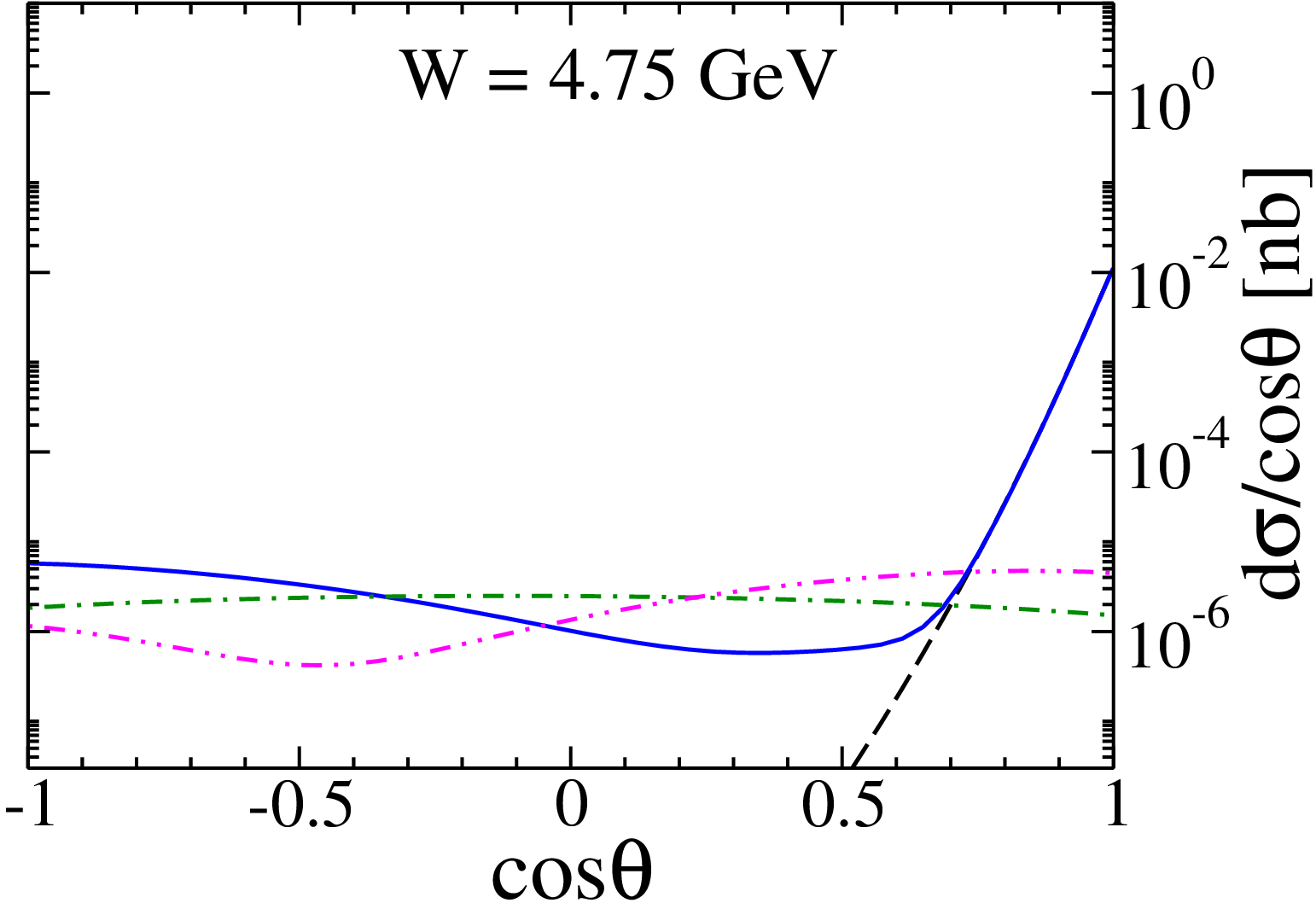}
\end{tabular}
\caption{Differential cross sections for $\pi^- p \to J/\psi n$ at
four different CM energies.
The spin-parity is selected as $J^P = (3/2^-,5/2^+)$ for 
($P_c^0(4380)$,\,$P_c^0(4450)$).}
\label{FIG7}
\end{figure}

In Fig.~\ref{FIG7}, we show the results of differential cross
sections for the $\pi^- p \to J/\psi n$ reaction at four different CM
energies $W$ near the threshold. As expected from the results obtained
for the total cross section, the contribution of the $P_c$ states has
a dominant influence on the differential cross section in the vicinity
of the energies corresponding to the $P_c$ resonances, so the
differential cross section is almost independent of the scattering
angle at $W=4.38$ GeV. As $W$ increases, the magnitude of the
differential cross section drops drastically. At $W=4.75$ GeV,
which is above the energies corresponding to the $P_c$ resonances, the
differential cross section exhibits a forward peak, 
where the dependence on $\theta$ is rather weak as $\theta$ increases. 

\section{Conclusion and summary}
\label{SecIV}
In the present study, we aimed to investigate the production mechanism
for the $\pi^- p \to J/\psi n$ reaction based on the hybridized Regge
model by including the contributions of the two heavy pentaquark
$P_c(4380)$ and $P_c(4450)$ states. First, we considered the effects of
$\rho$ and $\pi$ Reggeon exchanges, which yielded the background
contribution to the total cross section for the $\pi^- p \to J/\psi n$
reaction. The scale factor for the $\pi^- p\to \phi n$ was 
determined by fitting it to experimental data on the total cross
section. In order to avoid ambiguity, we employed the same form for the
scale factor to describe the $\pi^- p \to J/\psi n$ reaction by using the
same numerical values for the parameters used. Thus, we found that
the background contribution to the total cross section for the $\pi^-
p \to J/\psi n$ process is approximately $10^6$ times smaller than
that for the $\pi^- p\to \phi n$ reaction.  
This is mainly explained by the highly suppressed value of the coupling
constant $g_{J/\psi\rho\pi}$ compared with that of $g_{\phi\rho\pi}$ in the
case of dominant $\rho$-meson Reggeon exchange.

No information is available regarding the branching ratios of the
decays for heavy pentaquark resonances, so we carefully studied the
assumptions and suggestions proposed in previous theoretical
studies. First, we examined the branching ratios proposed by Wu et
al.~\cite{Wu:2010jy}, where we assumed that the predicted states in Wu
et al.~\cite{Wu:2010jy} correspond to the heavy pentaquark states
observed by the LHCb Collaboration. Wu et al.~\cite{Wu:2010jy}
proposed branching ratios for the pentaquark states of the $J/\psi N$
and $\pi N$ decay channels of about $40~\%$ and $6~\%$, respectively.  
Considering these values for the branching ratios, the 
results for the total cross section for $\pi^- p \to J/\psi n$ near 
the resonance region are $10^4$ nb, which is approximately of the 
same order as that for $\pi^- p \to \phi n$. 
However, if we use the branching ratio proposed by
Wang et al.~\cite{Wang:2015jsa}, the results obtained for the total
cross section decrease dramatically, where the maximum values are in
the order of 1 nb. This is consistent with the experimental upper 
limit~\cite{Jenkins:1977xb,Chiang:1986gn} on the total cross section
for the $\pi^- p \to J/\psi n$ reaction. The results are not sensitive
to the selection of different spins and parities for the heavy
pentaquark states. 

In the present study, we examined the two neutral pentaquark states 
$P_c^0$, but the other isosymmetric reaction $\pi^+ n \to
J/\psi p$ is as appropriate for the study of charged pentaquark states
$P_c^+$ as for the intermediate states. The total cross section of the
$\pi^+ n\to J/\psi p$ process is simply the same as that of $\pi^-
p\to J/\psi n$. However, for $J/\psi$ photoproduction,
we could have a large isospin asymmetry because of the different
photocouplings between $P_c^+$ and its neutral partner. We also
obtained the results for differential cross sections at four
different energies in the CM frame in the vicinity of the energies
corresponding to the heavy pentaquark resonance states.  

\section*{Acknowledgments}
S.H.K acknowledges support from the Young Scientist Training Program 
at the Asia Pacific Center for Theoretical Physics by the Korea Ministry
of Education, Science, and Technology, Gyeongsangbuk-Do and Pohang
City. H.-Ch.K. was supported by the Basic Science Research Program
through the National Research Foundation of Korea funded by the
Ministry of Education, Science, and Technology (Grant Number:
NRF-2015R1A2A2A04007048).  AH was supported in part by MEXT      
(Grant-in-Aid for Science Research (C) J26400273).



\end{document}